\documentclass[10pt]{article}
\usepackage{harvard}
\usepackage{graphics}
\usepackage{graphicx}

\begin{document}

\title{Utility Function and Optimum Consumption in the models with Habit Formation and Catching up with the Joneses}

\author{Roman Naryshkin$^{a,}$\thanks{Corresponding author, \texttt{rnaryshk@uwo.ca}, Tel: +1-519-661-3649, Fax: +1-519-661-3523}$\;$ and Matt Davison$^a$\\
$^a$\emph{Department of Applied Mathematics, University of Western Ontario} \\
\emph{1151 Richmond Str., London, ON, Canada}}
\maketitle

This paper analyzes popular time-nonseparable utility functions that describe ``habit formation'' consumer preferences comparing current consumption with the time averaged past consumption of the same individual and ``catching up with the Joneses'' (CuJ) models comparing individual consumption with a cross-sectional average consumption level.
Few of these models give reasonable optimum consumption time series. We introduce theoretically justified utility specifications leading to a plausible consumption behavior to show that habit formation preferences must be described by a power CRRA utility function different from the exponential CARA used for CuJ.

\bigskip


\bigskip

4.001  	    	    	 D1 - Household Behavior;\\
5.002  	    	    	 E2 - Consumption, Saving, Production, Employment, and Investment.

\newpage

\section{Introduction}

The main properties of the Morgenstern-von Neumann utility function $u(c_t)$ are convexity ($u''<0$) and time-separability meaning that utility is derived from the current consumption $c_t$ independent of the past. These simplifying assumptions have faced repeated theoretical criticism (\cite{KT1979,Shoemaker1982}) and lead to paradoxes when compared to the empirical data (\cite{Easterlin1974,MP1985}).

Numerous attempts to resolve these problems within the expected utility approach can be classified (following \cite{AMT2004}) into two categories by the ways in which the time-nonseparable utility function is constructed. The first, external criterion, category considers the averaged consumption of the economy as a standard of living benchmark compared to which the current individual's consumption preferences are computed. This ``catching up with the Joneses'' (CuJ) approach was introduced in \cite{Duesenberry1949} and developed in \cite{Abel1990} and \cite{Gali1994}. The utility function typically has the form $u(c/C^D),$ where $C$ is the per capita consumption in the economy (\cite{Gali1994,Abel1990}). The second, internal criterion, category uses individual's own past consumption as a benchmark for the current consumption. This ``habit formation'' approach was initially proposed in \cite{RH1973} and usually assumes that the current consumption $c_t$ is compared to the weighted aggregate $z_t=\int_{-\infty}^t e^{-a(t-\tau)} c_{\tau} d\tau$ or recent past $c_{t-1}$ consumption either additively ($u(c_t-bz_t)$) (\cite{Sundaresan1989,Constantinides1990,Smith2002}) or multiplicatively ($u(c_t/c^d_{t-1})$) (\cite{Abel1990,Fuhrer2000}) in the utility function. The papers \cite{Abel1990,AMT2004} and \cite{Gomez2007} attempt to combine both habit formation and CuJ in a single utility function.

The present work examines some existing time-nonseparable utility specifications and discuss their feasibility. Qualitative analysis is also used to construct utility functions which satisfy both analytical requirements and empirical data.

\section{Habit Formation}

\subsection{Critique of the short-memory utility function}

Consider a short-memory utility function in which the current consumption is compared to the past consumption during the last-period only, namely $u=u(c_t,c_{t-1}).$ The typical form used in the literature (e.g. \cite{Abel1990,Fuhrer2000}) is $u=u\left(\frac{c_t}{c^d_{t-1}}\right),$ where $d$ is a constant and $u$ has an isoelastic form $u_\mathrm{CRRA}(x)=x^\gamma/\gamma$ with a constant relative risk aversion $-xu''/u'=1-\gamma.$ To see whether such a utility function leads to a plausible optimum policy for consumption, we maximize the life-time discounted utility
\[ \max_{c_1, \dots, c_N} \sum_{t=1}^N e^{-\rho t} u_\mathrm{CRRA}\left(\frac{c_t}{c^d_{t-1}}\right) \]
subject to
\[ \sum_{t=1}^N c_t=W_0. \]
The direct optimization was carried out using the computer algebra software Maple 12, and the numerical results for risk aversion $1-\gamma=0.5,$ consumption period $N=20$ years, discount factor $\rho=3\%$, initial wealth $W_0=\$1,000,000$ and inherited consumption $c_0$=\$100,000 are shown in Fig. 1.
\begin{figure}
\includegraphics[angle=-90,width=\textwidth]{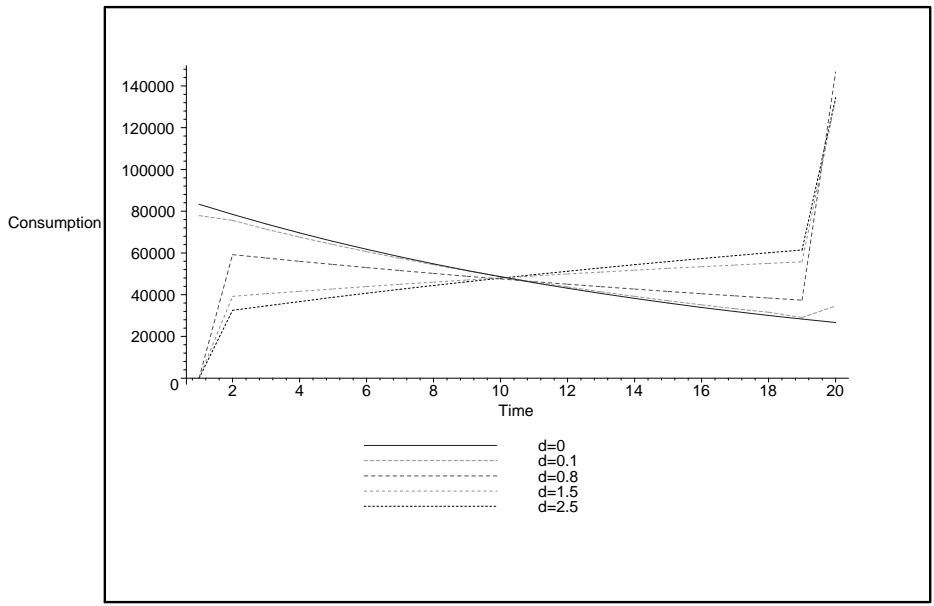}
\caption{Optimum consumption in a 20-year period under habit formation with short (1-year) memory.}
\end{figure}
The memory for one year only does make consumption grow with time for quite large $d,$ but also creates huge jumps in consumption at the first and at the last year. In a particular case of the logarithmic utility $u(x)=\log x$ (CRRA with $\gamma \to 0$) with $d=1$ and $\rho=0,$ the objective function simplifies to $$\max_{c_1, \dots, c_N} \sum_{t=1}^N \log \frac{c_t}{c_{t-1}}= \max_{c_1, \dots, c_N} \{ \log c_N - \log c_0 \},$$ and that it becomes optimal to wait until the end and consume all of the wealth in the last year, which is unrealistic.

A similar phenomenon is observed with an $M$-period utility for which the objective function takes the form ($d=1$):
\[ \max_{c_1, \dots, c_N} \sum_{t=1}^N  e^{-\rho t} u_{CRRA}\left(\frac{c_t}{c_0+\frac{1}{M}\sum\limits_{t'=t-M}^{t-1}c_{t'}}\right) \]
Numerical results for various memory lengths in Fig. 2 show
\begin{figure}
\includegraphics[angle=-90,width=\textwidth]{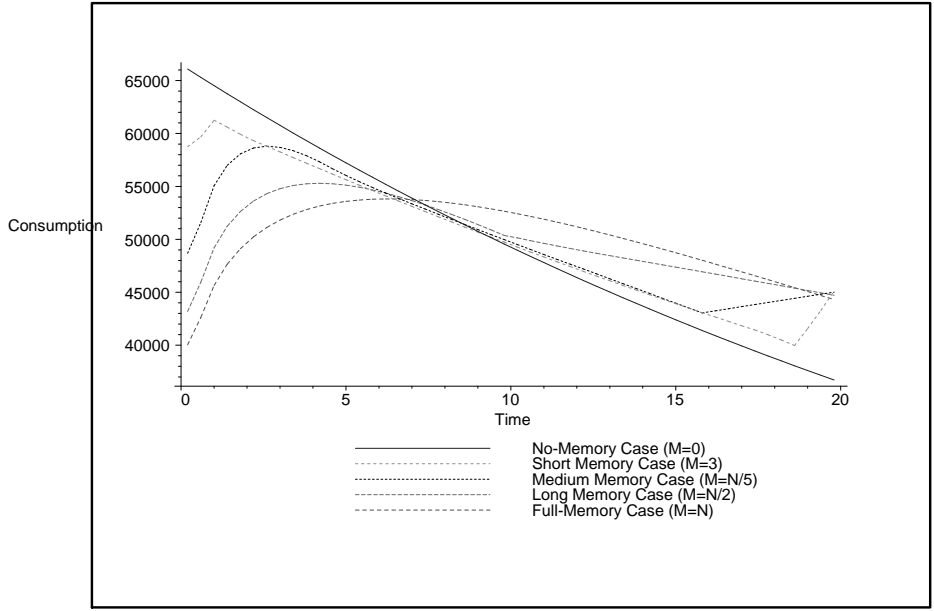}
\caption{Optimum consumption under habit formation with various lengths of memory.}
\end{figure}
that the length of the consumer's memory $M$ drastically changes his consumption in the first $M$ and last $M$ years, leaving intermediate consumption almost unchanged.

Avoiding abrupt changes in the optimum consumption behavior requires memory of every past period. Only in this case does the consumption become smooth and give plausible habit forming behavior  --- consumption increases as habits form, then becomes saturated and afterwards slowly decreases due to the discount factor $\rho$. Therefore there is a tradeoff between habit formation and the consumer's time preferences described by the discount factor.

The habit formation's strength can be varied by introducing a parameter $\beta$ such that the utility function becomes $u\left(\frac{c_t}{{c}_0+\beta \bar{c}_t}\right),$ where $\bar{c}_t$ is the averaged past consumption $\bar{c}_t=\frac{1}{t-1}\sum\limits_{t'=1}^{t-1}c_{t'}.$

\subsection{Critique of the additive utility function}

Another class of habit formation utility functions includes full memory about past consumption. Many authors (e.g. \cite{Sundaresan1989,Constantinides1990,Smith2002}) consider the utility of the additive form $u(c_t-b z_t),$ where $z_t=\int_{-\infty}^t e^{-a(t-\tau)} c_{\tau} d\tau$ is the weighted aggregate past consumption. For the isoelastic utility $u_\mathrm{CRRA}(x)=x^\gamma/\gamma,$ this specification implies that consumption $c_t$ has to stay above or equal to $b z_t.$ As a result of this ``additive=addictive'' utility, the optimum consumption suffers from the pathological solution allowing ``individual's bankruptcy'' $c_t=0$ for some $t>t_\mathrm{b}.$ This pathology remains, even if it is slightly reduced by replacing $z_t$ by the averaged consumption $\bar{c}_t=\bar{c}_0+\frac{1}{t}\int_{0}^t c_{\tau} d\tau,$ or its weighted analogue and simply stops working for quite high values of $b.$

One remedy might be to replace the CRRA utility by the constant absolute risk aversion (CARA) utility  $u_\mathrm{CARA}(x)=-e^{-\eta x}/\eta$ therefore allowing negative values of $x.$ %
\begin{figure}
\includegraphics[angle=-90,width=\textwidth]{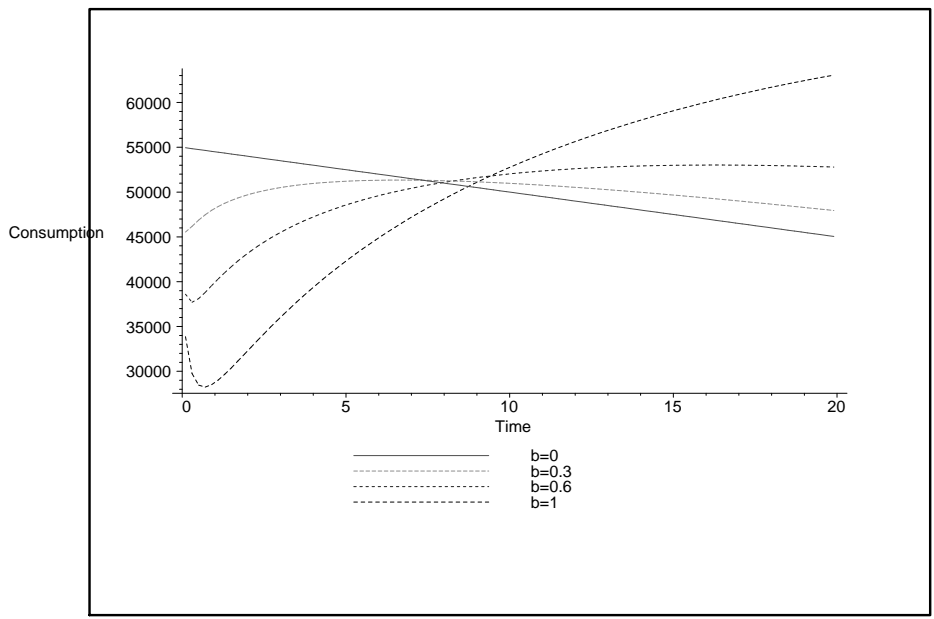}
\caption{Optimum consumption under habit formation with additive utility.}
\end{figure}
To see whether this is the case, consider the problem of optimizing the objective function
\[ \max_{c_1, \dots, c_N} \sum_{t=1}^N  e^{-\rho t} u_\mathrm{CARA}(c_t-b \bar{c}_t) \]
The optimum solution is shown in Fig. 3. This consumption path begins with an intuitively challenging initial trough. A better way to fix the problem was already mentioned in the previous section. The ``multiplicative'' utility $u(c_t,\bar{c}_t)=u\left(\frac{c_t}{\bar{c}_0+\beta \bar{c}_t}\right)$ is a good candidate for the habit formation both from a theoretical perspective as well as from the point of view of the optimum solution behavior. Numerical solutions for several parameters $\beta$ are shown in Fig. 4.
\begin{figure}
\includegraphics[angle=-90,width=\textwidth]{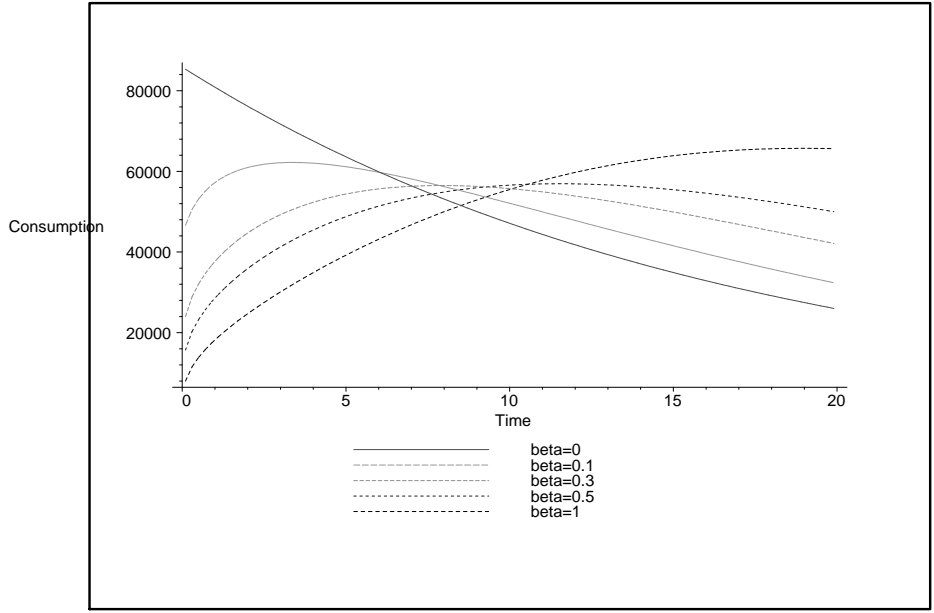}
\caption{Optimum consumption under habit formation with multiplicative utility.}
\end{figure}

\subsection{Utility function for habit formation}

Another type of utility function of the form $u(c_t,\bar{c}_t)=u_1(c_t)+\beta u_2(\bar{c}_t)$ was also examined and found to be not suitable for the description of habit formation because it led to decreasing consumption paths. In summary, we found that there are two simple types of utility functions that seem to give feasible optimum solutions for consumption:
\begin{itemize}

\item $$u(c_t,\bar{c}_t)=u_\mathrm{CRRA}\left(\frac{c_t}{\bar{c}_0+\beta \bar{c}_t}\right)$$

\item $$u(c_t,\bar{c}_t)=u_\mathrm{CRRA}\left(\frac{c_t}{\bar{c}^d_t}\right)$$

\end{itemize}

The latter form of utility is used by the authors \cite{AMT2004} and \cite{Gomez2007}. However, they also include a second part that represents CuJ and seems to be misused for that purpose, as explained in the next section.
%

\section{Catching Up with the Joneses}

\subsection{Critique of the multiplicative utility function}

The ``Catching up with the Joneses'' (CuJ) idea introduces external preferences into the consumer's behavior. An individual's level of consumption $c_t$ is directly related to the whole economy through the per-capita consumption $C_t$: if the economy grows, an individual will feel a need to increase his consumption to maintain his happiness. Therefore, one expects a positive change in the optimum consumption (relative to the standard time-separable solution) with a positive change in per-capita consumption. Unfortunately, some of the popular models of utility functions for CuJ lead, as we will see, to the opposite behavior.

Following \cite{Abel1990} and \cite{AMT2004,Gomez2007} we consider the utility function of the form
$$u(c_t,C_t)=u\left(\frac{c_t}{C^D_t} \right),$$ where $D$ is a tunable constant. We also assume that the per-capita consumption is a linear function approximately doubling every 30 years $C_t=C_0(1+t/30)$ (\cite{Gali1990}). The numerical results are shown in Fig. 5.
\begin{figure}
\includegraphics[angle=-90,width=\textwidth]{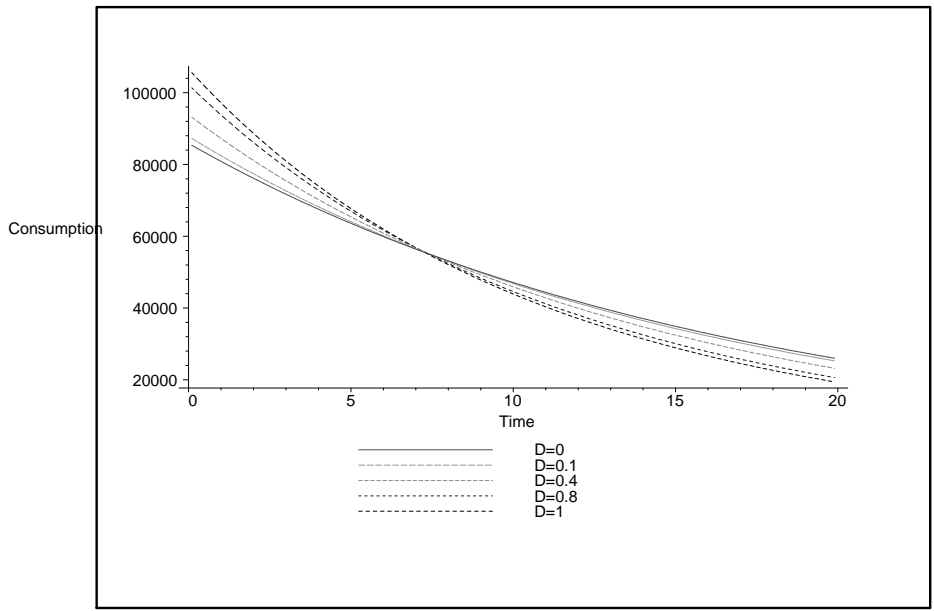}
\caption{Optimum consumption under catching up with the Joneses with multiplicative utility.}
\end{figure}
We can see that the consumption decreases even faster with time than in the standard model with $D=0$, opposite to the assumption of growing economy. This multiplicative utility type is therefore unsuitable for modeling CuJ. This result is easily extended to any growing function $C_t$ and is particularly obvious for exponentially increasing $C_t=C_0e^{\lambda t}$ which imply, for any positive $D$ and isoelastic preferences, decreasing in time utility $$u_\mathrm{CRRA}(c_t,C_t)=\frac{1}{\gamma}\frac{c_t^\gamma}{C_t^{\gamma D}}=\frac{1}{\gamma}\frac{c_t^\gamma}{C_0^{\gamma D}}e^{-\gamma\lambda D t}$$ equivalent to the utility without CuJ discounted, however, with a bigger factor $\rho'=\rho+\gamma\lambda D$. Moreover, even the case where $D$ is negative, equivalent to the utility function $u(c_t,C_t)=u(c_tC^{|D|}_t)$ considered by \cite{Gali1994}, fails to predict increasing consumption.

\subsection{Utility function for catching up with the Joneses}

A hint on modelling CuJ comes from the empirical ``Easterlin Paradox'' result (see \cite{Easterlin1974,CFS2008}) which shows that, despite sharp rises in per-capita GDP, average happiness has remained constant over time. Mathematically this behavior can be expressed as $u(c_t, C_t) = \mathrm{const}$ or $$\frac{du(c_t,C_t)}{dt}=\frac{\partial u}{\partial c}\frac{dc}{dt}+\frac{\partial u}{\partial C}\frac{dC}{dt}=0.$$ Empirical data suggest that $C_t=C_0(1+t/30)$ which gives ${dC}/{dt}=C_0/30=\mathrm{const},$ and assuming  $c_t$ to be approximately linear (confirmed by Fig. 6.) yields a simple partial differential equation of the form $$\frac{\partial u}{\partial c}+\alpha \frac{\partial u}{\partial C}=0,$$ with a constant $\alpha=\mathrm{const},$ which is easily integrated giving $$u=u(c_t-\alpha C_t).$$ So the utility function for CuJ must be additive.

Further numerical analysis shows that, similar to the case of habit formation, the additive utility cannot be assumed isoelastic (as in \cite{LU2000}) because of possible negative values of $c_t-\alpha C_t,$ so must be of the CARA type. In Fig. 6. we present optimal consumption for $$u=u_\mathrm{CARA}(c_t-\alpha C_t).$$
\begin{figure}
\includegraphics[angle=-90,width=\textwidth]{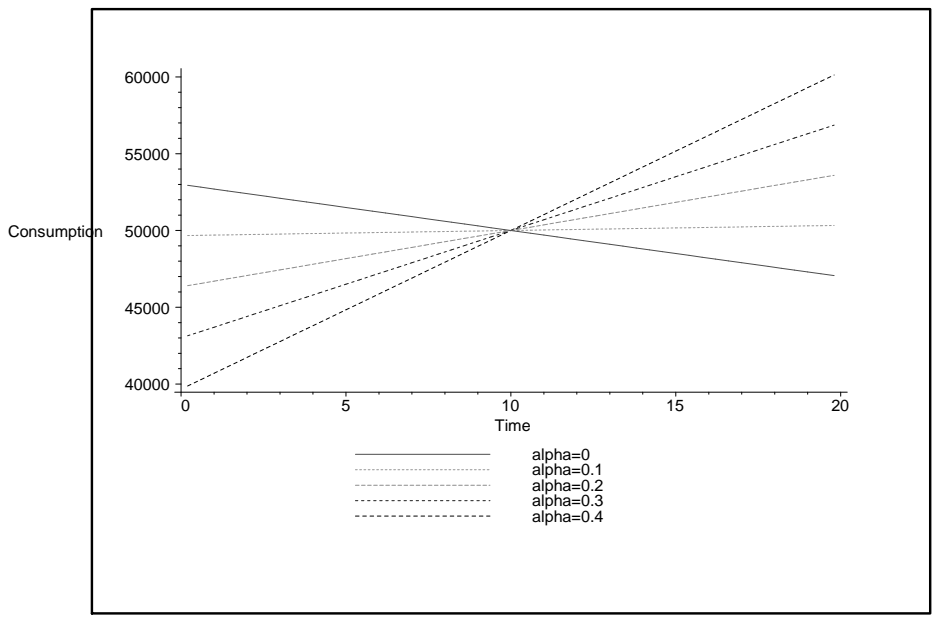}
\caption{Optimum consumption under catching up with the Joneses with additive CARA utility.}
\end{figure}

Two other types of utilities: $u\left(\frac{c_t}{C_0+\alpha C_t}\right),$ and $u_1(c_t)+\alpha u_2(C_t)$ were also considered and found unsuitable for CuJ because the former led to decreasing consumption paths and the latter didn't influence the optimum consumption at all ($C_t$ separates from $c_t$).

We conclude that empirical consumption growth evidence suggests a simple utility form which best describes consumption externalities should be exponential CARA type $u=u_\mathrm{CARA}(c_t-\alpha C_t).$

\section{Conclusions}

This paper examines several widely-used utility specifications designed to describe internal consumer preferences that specify her consumption path based on past consumption (habit formation), and preference externalities (catching up with the Joneses). Few of these models lead to feasible patterns in the consumption path of the consumer. We recommend how to change the utility function to correctly reflect the consumer's theoretical and empirical behavior.
\begin{figure}
\includegraphics[angle=-90,width=\textwidth]{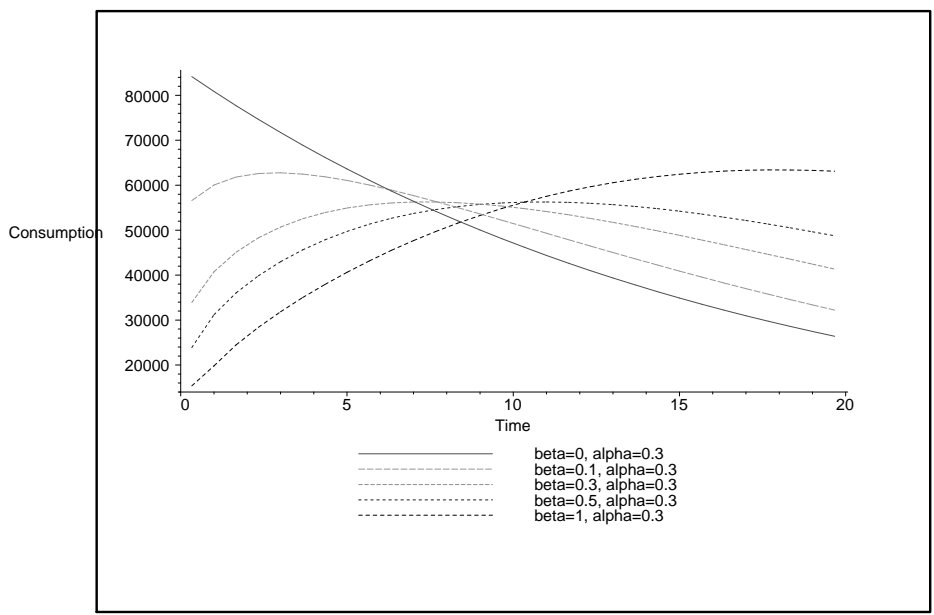}
\caption{Optimum consumption under habit formation and catching up with the Joneses.}
\end{figure}

The paper concludes that habit formation should be described by the isoelastic CRRA utility of the form $$u(c_t,\bar{c}_t)=u_\mathrm{CRRA}\left(\frac{c_t}{\bar{c}_0+\beta \bar{c}_t}\right)=\frac{1}{\gamma}\left(\frac{c_t}{\bar{c}_0+\beta \bar{c}_t}\right)^\gamma$$ and that catching up with the Joneses should be described by the exponential CARA utility $$u(c_t,C_t)=u_\mathrm{CARA}(c_t-\alpha C_t)=-\frac{1}{\eta}e^{-\eta (c_t-\alpha C_t)}.$$

These may be combined in a single utility function via the linear combination
$$u(c_t,\bar{c}_t,C_t)=Au_\mathrm{CRRA}\left(\frac{c_t}{\bar{c}_0+\beta \bar{c}_t}\right)+(1-A)u_\mathrm{CARA}(c_t-\alpha C_t).$$
The  $A=1$ corresponds to pure habit formation and $A=0$ corresponds to pure catching up with the Joneses. The optimum consumption for an intermediate case $A=1/2$ is shown in Fig. 7.

\end{document}